# AI for Biologists: The Knowledge Boundaries of Foundation Models in the Age of Agentic Systems

Tian Cai[1,*], Andrew Kleist[1], Jaimin Patel[1], Duccio Malinverni[1], Alexander K. Taylor[2], Malvika Sharan[3], Renato Umeton[3], and M. Madan Babu[1,3,*]

[1]Center of Excellence for Data-Driven Discovery, Department of Structural Biology, St. Jude Children's Research Hospital, Memphis, Tennessee 38105, USA

[2]Department of Computer Science, University of California, Los Angeles, California 90095, USA

[3]Office of Data Science, St. Jude Children's Research Hospital, Memphis, Tennessee 38105, USA

***Corresponding authors:** Tian Cai (tian.cai@stjude.org) and M. Madan Babu (madan.babu@stjude.org), St. Jude Children's Research Hospital, 262 Danny Thomas Place, Memphis, Tennessee 38105, USA

**ABSTRACT**

Artificial intelligence (AI) is changing how biological research is done. Biological foundation models trained on millions of nucleotides, protein, and single-cell measurements predict protein structures, mutation effects, and cellular responses to genetic perturbations. Agentic systems, AI assistants built on large language models, can run multistep analyses from plain-language requests and increasingly call foundation models as tools. Yet it is often unclear when these models' outputs can be trusted. We frame this as a model's knowledge boundary: the range of inputs, tasks, and conditions over which its outputs are reliable. This review explains how foundation models are built and proposes that five design elements (data, architecture, training scheme, evaluation scheme, and post-analysis) define this boundary. We examine evidence on where current models succeed or fail and present standardized descriptions of model limits so that researchers, and the agents acting for them, can use these tools reliably.

## 1. INTRODUCTION

Data science and computation have long been part of biological research, and their roles have deepened as researchers have pursued the goal of extracting knowledge from data: from processing raw data into signal, to interpreting that signal, and now to contributing to discovery itself. Computation initially served to process raw measurements, for example by performing the Fourier calculations that turned X-ray diffraction data into the first protein structure, myoglobin (1). As data grew in volume and variety, computation increasingly helped to interpret that signal: the Protein Data Bank, for example, grew from 7 structures at its founding in 1971 to more than 250,000 today (2, 3), and that growth drove new methods for comparing and classifying structures. New disciplines formed around these tasks, including biostatistics, bioinformatics, and computational biology. AI is now contributing to discovery directly: models that learn from data can predict outcomes or propose hypotheses not yet tested experimentally, as AlphaFold does when it predicts protein structure from amino acid sequence (4, 5). Two classes of complementary AI tools have extended this shift across biology (**Figure 1A**): (i) biological foundation models, trained on millions of sequences, protein structures, or cellular measurements, offer versatility: general patterns are extracted from data once, when a model is developed, and can then be applied to address many additional biological problems; and (ii) agentic systems, which run analyses from plain-language instructions, offer accessibility: biologists with limited programming experience can now apply advanced computational tools to their own datasets. Together, these tools have drawn biologists and AI researchers into close collaboration, and AI for biology is emerging as a distinct interdisciplinary field. Because this convergence is new to both fields, we do not have a clear picture of what these models have learned and where they fail.

Foundation models and agentic systems extract knowledge in different ways (**Figure 1B**). Foundation models compress large bodies of raw data into richly organized numeric representations. Protein language models such as ESM2 are trained on UniRef sequences (6), genomic models such as Nucleotide Transformer and Evo 2 learn from genomes spanning diverse species (7, 8), and single-cell models such as Geneformer, scGPT, and UCE learn from measurements made on tens of millions of cells (9-11). What a foundation model 'knows' is therefore bounded by what its training data contain and by how much of that content its architecture and training objective allow it to capture. Agentic systems compose tools into workflows: an agent uses a large language model (LLM) as its source of reasoning to plan the specific steps (i.e., tasks) and carries them out by assembling literature, databases, and software. Agents run tools on raw data, but their choices of which tools to run and how to interpret the results draw on the LLM's knowledge, which is restricted to what has previously been documented. Their recognized successes have accordingly been in the space of conventional bioinformatics and biostatistics (12-14), although independent evaluations show that current agents remain brittle on realistic analysis tasks (15, 16). Together, the two turn raw biological data (e.g., a protein sequence) into a reusable intermediate form (e.g., a sequence *embedding*) and deliver it to diverse downstream applications and users (**Figure 1C**).

For instance, AlphaFold, a protein structure prediction model, and Biomni, an agentic bioinformatics system, both draw on public protein resources such as UniProt and the Protein Data Bank (4, 17). AlphaFold can accurately predict the structure of an uncharacterized protein from its sequence, which an agent composing existing software cannot do without AlphaFold or a related model. Once a structure is predicted, however, many further analyses, such as comparing it with known structures, relating it to variants or expression data, and planning follow-up experiments, are the kind of multistep work an agent can assemble. The two are beginning to meet. Agentic systems such as Biomni and the Virtual Biotech now call foundation models as tools (17, 18), and multimodal models such as ChatNT, CellWhisperer, and BioReason feed foundation-model embeddings directly into an LLM (19-21). Yet it remains unclear how an agent, or a researcher, should choose the right foundation model for a task and how to interpret its output.

What limits the broader use of both foundation models and agentic systems in biology is the lack of an explicit account of each model's knowledge boundary: the range of inputs, tasks, and conditions over which its outputs can be relied upon. Such an account would (i) help researchers decide when a model can be trusted and how to evaluate its output and (ii) give agentic systems the information they need to use foundation models as reliable components of larger workflows. This review discusses the conceptual foundation of knowledge boundaries. We first introduce the design elements that underlie biological foundation models and interpret existing models within this framework. We next describe how agentic systems are assembled and the routes by which they interface with foundation models. We then review the evidence, including independent reassessments against simple baselines, on what these models capture and what they miss, tracing both back to their design elements. Finally, we present practical ways to make knowledge boundaries explicit to both researchers and agents. We focus on molecular sequences, macromolecular structures, single-cell data, and other omics data, although we envision that the concepts apply more broadly in biology and medicine. Key terms used in this article are defined in **Supplemental Table S1**. An online resource accompanying this review (**HTML file**) provides an interactive landscape of the various biological foundation models discussed here.

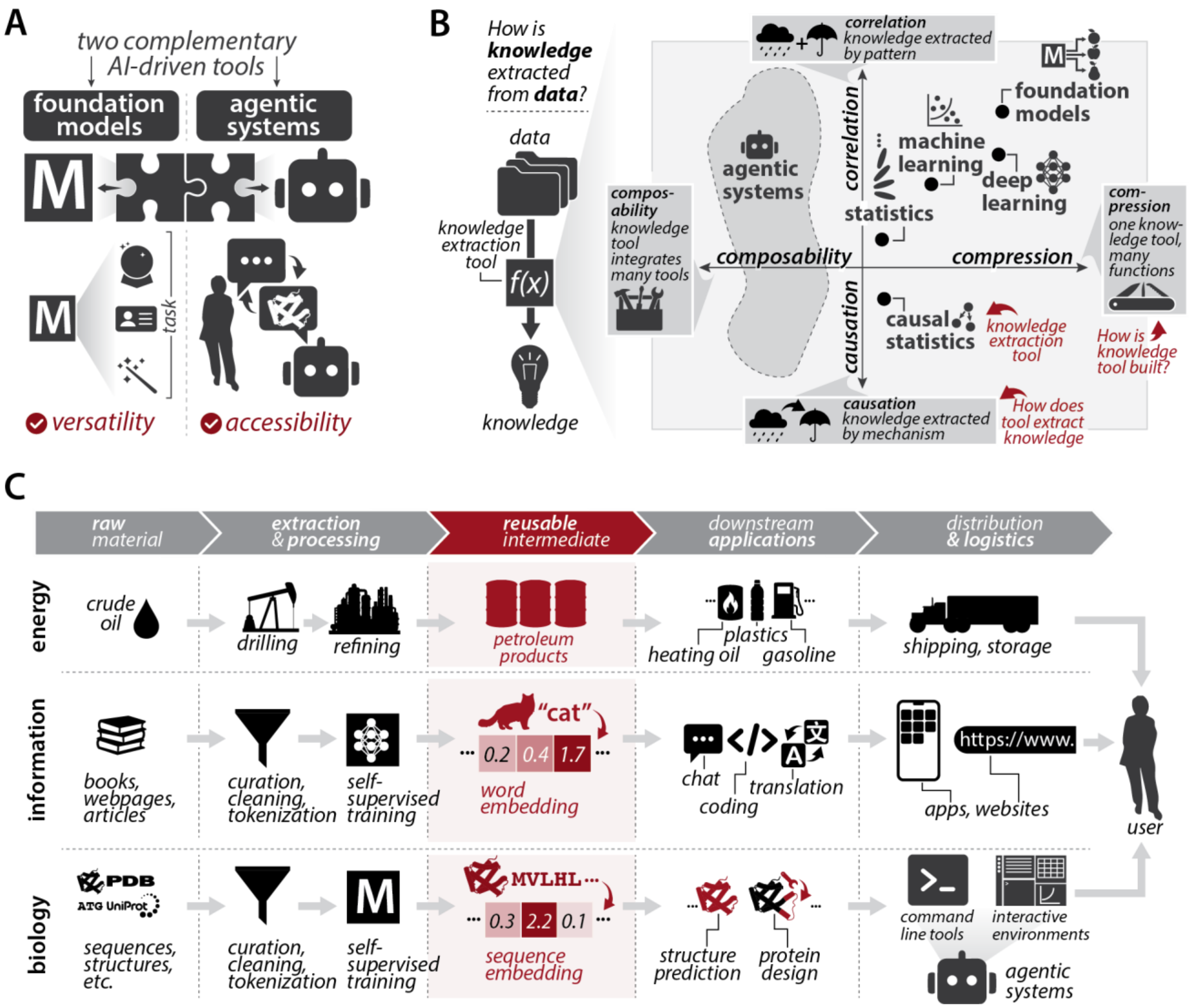

**Figure 1. Complementary roles of biological foundation models and agentic systems. A.** Foundation models and agentic systems serve complementary functions. **B.** Different computational and statistical tools extract knowledge from data, and these tools differ from one another in (i) how they are constructed (e.g., composability vs. compression) and (ii) how they extract knowledge (correlation vs. causation). Foundation models develop higher-order 'knowledge' from data by compressing data into dense numeric mathematical objects called embeddings. Their knowledge develops as a consequence of sophisticated computational approaches for pattern recognition. As such, foundation models can make accurate predictions (e.g., predictions of protein structure from amino acid sequence), without having 'learned' the mechanisms driving the phenomenon they predict (e.g., physical and chemical basis of protein folding). Agentic systems function to integrate and link tools (including foundation models), literature, databases, and data into workflows. Because they function primarily as integrators of existing tools, they take on the properties of the tools they utilize in any specific instance. **C.** Extraction, processing, and distribution of raw data in the form of a reusable intermediate by analogy to crude oil processing.

## 2. LANDSCAPE OF BIOLOGICAL FOUNDATION MODELS

### 2.1. What Is a Foundation Model and What Is Its Premise?

The term "foundation model" was introduced in 2021 to describe models trained on broad data that could be adapted to a wide range of downstream tasks (22). A common way of training a foundation model is self-supervised learning. During training, part of each training example is hidden ("masked"), and the model is asked to predict the hidden part from the parts that remain visible. For example, a few amino acids in a protein sequence might be masked; the model predicts which amino acids belong at those positions, its predictions are compared with the true amino acids, and the model is adjusted to reduce its errors. This process is repeated across millions of sequences. Unlike supervised learning, in which humans must annotate ("supervise") the data, self-supervised learning creates its training targets from the raw data itself. This suits biology, where unlabeled DNA sequences, protein sequences, and single-cell transcriptomes number in the hundreds of millions, whereas validated labels, such as a variant classified as pathogenic or benign, are scarce, labor-intensive, and unevenly distributed. We also discuss models trained for a specific task that are not typically called foundation models, including AlphaFold for protein structure prediction (4, 5), AlphaGenome for sequence-to-function prediction (23), and perturbation models termed virtual cells (24), because they rest on the same premise and are frequently used as foundation models.

The premise of a foundation model is that it can be trained once on large unlabeled datasets and then reused for many different tasks. Training is designed to make the model extract recurring patterns in its data type: for proteins, pairs of amino acids that tend to mutate together because they contact each other in the folded structure; for DNA, short sequence motifs bound by transcription factors, such as the TATA box found in many promoters; for single-cell transcriptomes, groups of genes whose expression rises and falls together because they act in the same pathway. Once trained, the model can be reused in three ways. In probing, the model's numerical summary of each input (its embedding, defined below) is passed to a separate, much simpler model, such as a linear regression, which is trained on a small, labeled dataset to predict a property of interest. In fine-tuning, the foundation model itself is trained further on the small, labeled dataset. In zero-shot prediction, no labeled data are used at all. For example, a protein language model can score how plausible a sequence is, given the sequences it was trained on; a mutation that makes a sequence much less plausible is more likely to disrupt protein function, so this score can serve as an estimate of variant effect (25).

#### *2.1.1. Key Concepts to Understand Biological Foundation Models*

Biological foundation models borrow both their architecture and much of their terminology from large language models (LLMs). Like any computational tool, a foundation model takes an input, transforms it, and produces an output. Four concepts describe this process: the units a model reads (vocabulary and tokens), the rules by which those units combine (grammar), the data on which the model is trained (corpus), and the numerical representation it produces for each input (embedding). We introduce each concept through the analogy to human language, because the points at which the analogy breaks down foreshadow the knowledge boundaries discussed below.

**Vocabulary and token.** A model's vocabulary is the set of basic units it can recognize, analogous to the words of a language. A token is one such unit as it occurs in a particular input, so that an input is read as a sequence of tokens, much as a sentence is read as a sequence of words. General-purpose LLMs break text into tokens that correspond to words or word fragments. In biological foundation models, the equivalent tokens are amino acids, nucleotides, or short sequence motifs; in single-cell models, they are genes, with each cell read as a list of the genes it expresses. In ESM2, for example, the 20 standard amino acids comprise the model's vocabulary, so a protein of 300 amino acids is read as 300 tokens (6). Unlike human language, in which spaces and punctuation demarcate words, biological sequences such as DNA lack analogous boundaries, so model builders must choose among single nucleotides, fixed-length *k*-mers, or learned multi-nucleotide tokens (26, 27), and the tokenization rule becomes part of the scientific problem itself.

**Grammar.** As in natural language, the grammar of a biological language is the set of rules by which its units are combined. For proteins, these rules are dictated by evolution: the gene encoding an amino acid sequence persists only if the translated protein folds into the required structure and adopts the right conformational dynamics to perform its function. For regulatory DNA, the noncoding sequences such as promoters and enhancers that control when, where, and how strongly a gene is transcribed, the basic units are transcription factor (TF) binding motifs: short sequences, typically 6 to 12 base pairs long, each preferentially bound by a particular TF. The grammar lies in how motifs are arranged: the number, spacing, orientation, and order of motifs within a regulatory region determine which TFs bind together and, in turn, how strongly the target gene is expressed and in which cell types. Biological foundation models, trained to predict missing or next tokens across many contexts, extract patterns that reflect some of these rules (Section 4.2). Whereas an LLM implements grammar that humans devised and that any fluent reader can check, biological foundation models implement a grammar whose rules are largely unknown. When such a model performs well on a test, it can therefore be difficult to assess whether it has internalized general principles or simply found a shortcut that happens to work on that test.

**Corpus.** AI models rely on large repositories of training data termed corpora. Examples include UniRef, which contains hundreds of millions of protein sequences and was used to train ESM2 (6, 28), and the CZ CELLxGENE Discover database, which contains single-cell measurements from over 150 million cells and supplied the training data for scGPT (10, 29). These collections are substantial but remain smaller than text corpora: ESM3, among the largest protein models, was trained on about 771 billion tokens (30), roughly 20-fold fewer than the 15 trillion tokens used to pretrain the LLM Llama 3 (31), although tokens from the two domains are not directly equivalent in information content. Their effective diversity is also lower than their size suggests. Many sequences are near-duplicates of close evolutionary relatives, which is why protein databases are clustered by sequence similarity before training (28), and sequences are concentrated in a few well-studied lineages, which biases model likelihoods toward those lineages (32). Text corpora are also incomplete, but the incompleteness of the biological corpus differs in kind: whereas text records language directly, biological data are indirect measurements of an underlying system, shaped by assay technology and experimental design. Biological corpora therefore contain only what current methods can measure and what researchers have chosen to study, and consequently they inherit biases in species coverage, cell or tissue sampling, and assay type.

**Embedding**. An embedding is a long list of numbers, typically in the hundreds to thousands, that a model assigns to each token of a given query. It represents intrinsic properties of that token (e.g., amino acid size and charge) as well as contextual features arising from its relation to the other tokens in the input. Because embeddings are derived from training without human-interpretable labels, they cannot be read or understood directly: a human could not identify which number reflects, say, amino acid polarity, and individual numbers may represent a combination of properties. One strategy to address this problem is to train an auxiliary model, termed a sparse autoencoder, that re-expresses each embedding as a combination of a much larger set of features, each of which tends to correspond to a single property. The aim recalls the "Jennifer Aniston neuron" of neuroscience, a single neuron in the human brain that fires selectively to images of one person (33); we return to what such analyses have revealed in Section 4.2. Because all embeddings from a given model share the same numeric space, inputs with similar properties lie close together. For example, in ESM2, proteins with similar structure or function lie near one another even when their sequences differ (6).

Taken together, these considerations suggest that learning biology from biological data is closer to deciphering an extinct language than to reading a living one: the vocabulary must be inferred, no living speaker can confirm the grammar, and the only surviving texts are the measurements our assays have recorded. Section 4 uses this analogy to organize the evidence.

#### *2.1.2. Five Design Elements of Typical Foundation Models*

Much of the value of a foundation model lies in its ability to generalize, that is, to make useful predictions for inputs it did not encounter during training. ESM2, for example, can score a mutation in a protein that was absent from its training data. This ability has limits, however, and they are set by how the model was built: ESM2 cannot score nucleotide substitutions, because its vocabulary contains only amino acids. Although biological foundation models have multiplied rapidly over the past five years, nearly all of them can be described by five common design elements (**Figure 2A**), which we illustrate with ESM2 (6). Innovations in biological foundation models often result from changes to one or more of these elements, and each element also sets part of a model's knowledge boundary (Section 4.1).

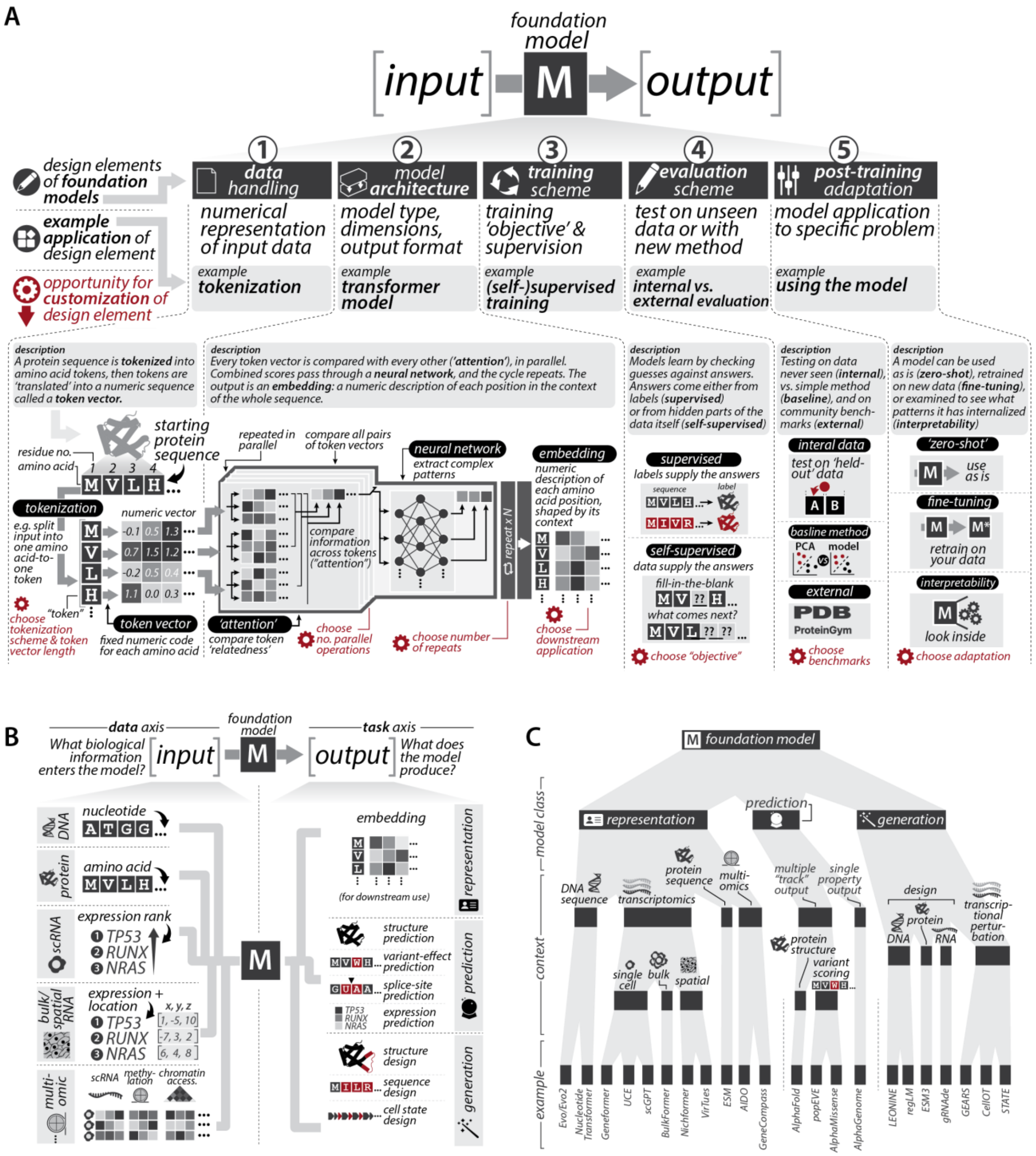

**Figure 2. Design elements of biological foundation models. A.** Design elements influence the performance, application, and limitations of foundation models. For each design element, an example and description is given, with key terms highlighted. Opportunities for those building or using foundation models to customize each element are shown in red. **B.** Different inputs (data

axis) and outputs (task axis) are highlighted. **C.** Hierarchical organization of different biological foundation models considered in this manuscript.

**Data** determines what information goes into the model and in which form, including the training corpus and tokenization strategy. ESM2 is trained on UniRef protein sequences tokenized as individual amino acids. Geneformer, by contrast, encodes each cell as a list of genes ranked by expression (9).

**Model architecture** determines how information is processed to extract and store learned properties, such as protein grammar. The leading architecture, used by ESM2, is the Transformer, whose self-attention relates every token to every other token in the input (34). Transformers carry few modality-specific assumptions (inductive biases), which enables them to adapt to very different corpora. However, this means they generally require larger training sets to learn modality-specific patterns. Alternative architectures instead build on features of the data, such as long genomic context (e.g., HyenaDNA) (35), the equivalence of the two DNA strands (e.g., Caduceus) (36), or protein interaction networks (e.g., PINNACLE) (37).

**Training scheme** determines what the model is rewarded to learn, and there are two popular directions. The first hides part of the input and asks the model to fill it in. In masked language modeling, used by ESM2, randomly hidden amino acids are predicted from the surrounding sequence (38). In autoregressive modeling, used by ProGen2 and Evo, each token is predicted from those before it (39-41), which makes these models natural generators of new sequences. The second direction teaches the model which samples belong together. Contrastive and metric learning pull related examples close together in embedding space and push unrelated ones apart. Orthrus, for example, treats splice isoforms and orthologs of the same gene as related RNAs (42), and SCimilarity places cells of the same type near one another (43).

**Evaluation scheme** determines how the model is judged, and it operates at two levels. Intrinsic evaluation asks whether the model has learned the structure of its data, for example whether ESM2 recovers residue contacts (6). Downstream evaluation asks whether the model helps with a practical task, such as predicting the effects of mutations, and is increasingly done on shared benchmarks such as ProteinGym (44). The two do not always agree.

**Post-analysis** determines how a trained model is used. Its central idea is transfer learning: knowledge captured in the embedding during pretraining is carried over to a new task by probing or fine-tuning (Section 2.1), so that a small, labeled dataset suffices where training from scratch would require a large one. A model can also be used as it is, with no further training, for example to score mutations by likelihood or, as Geneformer does, to simulate deleting a gene (*in silico* perturbation) and see how a cell's embedding shifts (9). Finally, interpretability analysis asks how the model arrives at its outputs, for example which input positions drive a prediction (45) or which internal features match known biology, and can reveal whether a model relies on genuine biology or on shortcuts in the data.

## 2.2. An Overview of the Biological Foundation Models

Biological foundation models can be differentiated by their inputs, the type of biological data that enters a model, and by their outputs, the task the model performs (**Figure 2B**). Models differ in their inputs mainly in how they define vocabulary, assemble a corpus, and choose an architecture suited to each data type, and in their outputs mainly in training scheme and post-analysis. Whereas a general-purpose LLM serves many kinds of text and tasks, each biological foundation model is shaped by design choices specific to its data and task and is therefore foundational mainly within that range. This dependence is the basis of the knowledge boundary developed below.

### *2.2.1. Inputs: Biological Data Types*

Moving from DNA and protein to cells and tissues, the unit of vocabulary grows from a single residue to a gene, a whole cell, and a cell within its neighborhood (**Figure 2B**), and the learnable grammar changes accordingly.

**DNA sequence.** For DNA language models, the first design question is how to define a token. Nucleotide Transformer adopts fixed six-nucleotide *k*-mers (7), whereas DNABERT-2 and GROVER learn variable-length tokens from genomic sequences (26, 27). Such multi-nucleotide tokens can split regulatory motifs across tokens and hide the effect of a single-nucleotide variant. To retain single-nucleotide resolution over long distances, HyenaDNA, Caduceus, and Evo keep one nucleotide per token and use architectures that remain efficient over long inputs (35, 36, 41).

**RNA sequence.** RNA language models read the nucleotide sequences of RNA molecules, unlike the single-cell models described below, which read how much of each RNA a cell contains. Because RNA is single-stranded, it can fold back on itself to form structures that are central to its function. Models such as RNA-FM and RiNALMo, trained on tens of millions of noncoding RNAs, recover which bases pair in the folded structure, much as ESM2 recovers protein contacts (46, 47). For mRNA, whose function depends on coding sequence, splicing, and regulatory regions, Orthrus instead uses contrastive learning across splice isoforms and orthologs (42). A model built for noncoding RNA structures therefore need not capture what governs mRNA function.

**Single-cell transcriptomics.** Single-cell models also describe RNA, but their token is a gene rather than a nucleotide. This choice already carries a cost that biologists will recognize: because expression is summarized at the gene level, transcript isoforms are not represented.

Their corpus is a matrix of expression counts across various cells, tissues, or cell lines, and their grammar is co-expression patterns. A central challenge has been how to represent expression levels, because a cell has no inherent order and each gene carries a continuous value rather than a discrete identity. Furthermore, platform- and processing-specific details can generate strong nonbiological confounding factors in gene expression datasets. To address this, Geneformer orders genes by expression rank, so that expression is encoded only by position in the input and not its quantitative value (9). By contrast, scGPT pairs each gene with a binned expression value, representing magnitude explicitly (10). These models share the same limitation: their vocabulary is a fixed list of gene identities from specific genome annotations, restricting them to the species seen in training. This is a limitation that UCE addresses (11).

**Bulk and spatial omics.** Bulk RNA sequencing measures the average expression of all cells in a tissue sample. It loses cellular resolution, but decades of bulk profiles in public archives come linked to tissue type, disease, and clinical outcome. BulkFormer is trained on such samples, keeping the gene as the unit of vocabulary while learning tissue-level patterns (48). Spatial transcriptomics instead records where each cell sits in a tissue section, so the grammar extends beyond which genes are co-expressed within a cell to which cells sit next to one another. Nicheformer is trained jointly on dissociated and spatial data, so that its embedding of a cell reflects both its own expression and its tissue neighborhood (49), and VirTues extends spatial modeling beyond RNA to protein markers in multiplexed tissue images (50).

**Protein sequence.** Protein foundation models, most of them protein language models, learn from amino acid sequences, which fold into the structures through which proteins carry out most cellular functions. The 20 natural amino acids form a natural vocabulary. ESM2 and ProtTrans established how much can be learned from sequence alone at scale (6, 51). Because the vocabulary is settled, later models have innovated in other design elements: SaProt and ESM3 add structure and function tokens (30, 52), ProGen2 adopts next-token prediction for generation (40), and Ankh optimizes architecture and training to match larger models at a fraction of the pretraining cost (53).

**Multi-omics.** Multi-omics foundation models learn from more than one molecular layer, such as DNA, RNA, protein, and chromatin, to represent how these layers jointly shape cell state. Combined modalities lack a shared vocabulary, and the central challenge is aligning a nucleotide, a gene count, and a protein structure in one space. Models take different routes around this challenge. GeneCompass enriches gene tokens with prior knowledge (54). PINNACLE starts from protein interaction networks built separately for each cell type from single-cell data and uses a graph neural network, which updates the embedding of each protein with information from its interaction partners*,* so that the same protein receives a different embedding in each cell type (37). AIDO instead builds separate DNA, RNA, protein, and cell models and combines their embeddings (55).

### *2.2.2. Outputs: AI Tasks*

Foundation models also differ in what they produce. We group their outputs into three types of tasks: representation, prediction, and generation (**Figure 2B,C**).

**Representation.** A representation model converts each input into an embedding. Because it is trained without labels (self-supervised training), it can draw on very large unannotated corpora, and the resulting embedding is a general-purpose resource: ESM2 embeddings, for example, have been used to predict structure, annotate function, and score mutations. What a model embeds, whether a residue or protein (ESM2), a DNA segment (Nucleotide Transformer,

Evo), a gene in its cellular context (Geneformer, scGPT), or a whole cell (UCE), determines which questions the embedding can serve (6, 7, 9-11, 41, 46).

**Prediction.** A predictive model is trained on paired measurements (supervised training) to answer one fixed question, so it can predict only what has been measured. AlphaFold takes an amino acid sequence and returns three-dimensional coordinates for every atom of an amino acid, based on structures in the Protein Data Bank (4). For DNA sequences, AlphaGenome takes up to one megabase of sequence and outputs numerous tracks, such as RNA expression, chromatin accessibility, and transcription factor binding, at up to single-base resolution (23), extending earlier models such as Enformer and Borzoi (56, 57). ChromBPNet takes the opposite approach, modeling a single assay, chromatin accessibility, at base resolution while separating nuclease cleavage bias from the underlying regulatory signal (45). Scooby adapts Borzoi to predict these signals in individual cells (58). Variant-effect predictors take a nucleotide or amino acid change and score how likely it is to cause disease. Because clinical labels are scarce, they rely on other signals: EVE uses evolutionary conservation (59), AlphaMissense adapts AlphaFold with human population frequencies (60), and popEVE combines both to make scores comparable across proteins (61).

**Generation.** A generative model writes new sequences or states rather than scoring existing ones, and its vocabulary determines what it can write. ProGen takes a protein-family tag and writes a full protein sequence one amino acid at a time (40, 62). ESM3 can instead be prompted with partial structure or function (30). For nucleic acids, generation spans genomes (Evo), regulatory DNA with specified activity (regLM), organism-specific codons (CodonTransformer), and RNA that folds into a target structure (gRNAde) (8, 41, 63-65). Perturbation models lie between prediction and generation: given a cell and a perturbation (e.g., a gene knockout), they produce the expected expression profile, a state that was never measured. Because sequencing destroys the cell, the same cell cannot be profiled both before and after a perturbation; only separate populations of untreated (control) and perturbed cells are available. CellOT learns how each control cell would most plausibly change, pairing it with a matching cell in the perturbed population, and can then predict the response of new cells (66). GEARS uses a gene-relationship graph to handle unseen perturbation combinations (67), and State, trained on a large perturbation corpus, treats whole cells as tokens (24).

## 3. WHERE A FOUNDATION MODEL ENTERS AGENTIC SYSTEMS

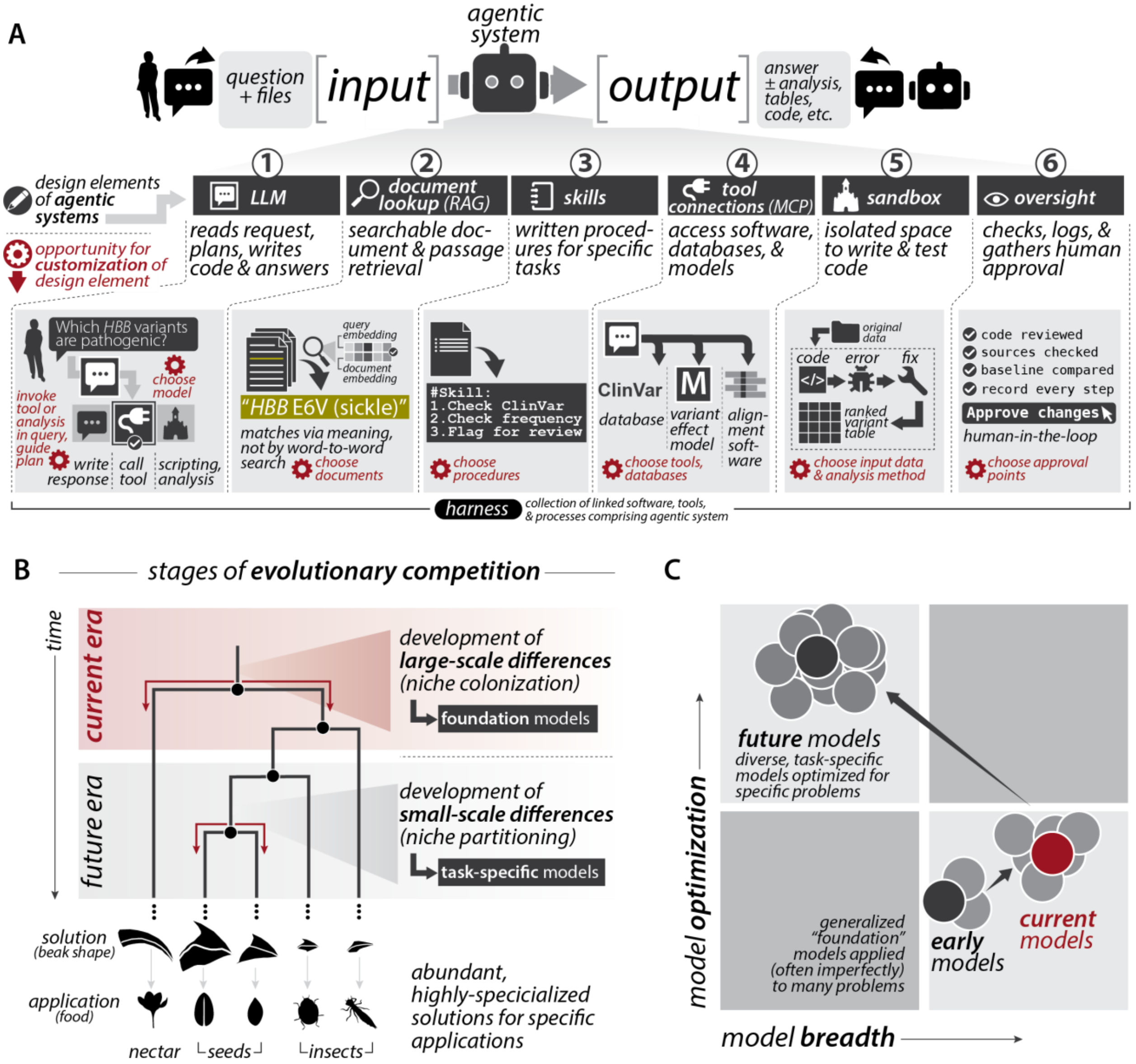

**Figure 3. Design elements of agentic systems and developmental maturity of AI-based tools for biology. A.** Design elements influence the performance, application, and limitations of agentic systems. Opportunities for those building or using agentic systems to customize each design element are shown in red. **B.** Early evolutionary specialization favors large-scale differences (niche colonization, analogous to current foundation models), whereas later adaptation produces small-scale differences optimized for specific niches (analogous to task-specific models). **C.** We anticipate that future models will be less broadly applicable but substantially better optimized for specific tasks, thereby filling in boundary regions for which current general models are less well optimized.

Biologists increasingly reach foundation models through agents. An agent uses a general-purpose LLM, such as those behind ChatGPT or Claude, to carry out a multistep task: asked which genes respond to a drug in a single-cell dataset, it can plan the analysis, write and run code, call a foundation model, check the results, and return to the user when a decision is needed (**Figure 3A**). Agentic systems are therefore not models themselves; they coordinate an LLM, the user, and many tools. Their components arose in overlapping waves, each a response to a limitation exposed by the one before. Chat-based LLMs, widely available from late 2022, brought these models into everyday scientific use (68) and revealed their limits: a chat model draws only on text fixed at a training cutoff, cannot say where an answer came from, sometimes produces fluent but false statements (hallucinations), and cannot act on data or run software.

The first responses connected the model to outside information. (a) Retrieval-augmented generation (RAG) has the LLM model generate its answer based on retrieved information from documents, such as papers or database entries, so that answers are current and traceable to a source (69). For instance, MyGPT is an open-source RAG platform that uses open-source LLMs, runs locally, or at remove facilities without internet access or within secure environments so that confidential and sensitive data never leaves the computer. Moreover, it provides confidence metrics including a hallucination index and provide explicit source citation from the library to assess contextual support for generated answers (70). (b) Function calling lets the model write a request to an external program, such as a database query or an analysis script, and read back the result (71, 72). (c) Reasoning models, such as OpenAI's o-series, DeepSeek-R1, and the "thinking" modes of Claude and Gemini, are LLMs trained to write out intermediate steps and rewarded when the final answer to a checkable problem is correct (73); this makes multistep answers more reliable but adds no information that was absent from training (74). Only external components, such as documents, tools, or a connected foundation model, can extend what the system draws on.

As tools multiplied, the Model Context Protocol (MCP), released in late 2024, standardized how they connect to LLM-based agents, much as USB did for computer devices: a database, program, or foundation model wrapped once as an MCP tool can be used by any compatible agent (75). Because access to a tool does not teach an agent to use it well, skill files followed in 2025: written instructions an agent loads when needed, such as the steps of a standard analysis or when a tool's output should not be trusted (76). Agent harnesses manage long-running tasks by repeating the plan–act–revise cycle, providing a workspace for codes (the sandbox in **Figure 3A**) and files, and asking a human for approval at key decisions (77).

Because these components are general-purpose, biological agents are assembled quickly. Biomni composes workflows from tools and protocols mined from thousands of publications (17), and Paper2Agent turns a paper's codebase into MCP tools (78). Other agents target chemistry, gene-set annotation, gene editing, single-cell reanalysis, and therapeutics (12-14, 79, 80). Multi-agent systems assign roles to separate LLM agents: in the Virtual Lab, agents acting as a principal investigator and specialist scientists designed nanobodies, and the AI co-scientist and Robin extend this to hypothesis generation (81-83).

A biological foundation model can be combined with an LLM-based agent (built on a general model such as GPT or Claude) in four ways. They differ in what the LLM-based agent receives.

**(i) Biological foundation model as a tool.** The agent calls the foundation model like any other software tool and reads its output, such as a predicted structure or score. For example, the Virtual Lab agents called ESM, AlphaFold-Multimer, and Rosetta to refine nanobodies (81). Paper2Agent's AlphaGenome tools outperformed general agents, which shows that how a model is packaged as a tool matters (78, 84).

**(ii) Molecular measurements as training data.** Molecular measurements are written as text and used as training data to fine-tune the LLM-based agent. For example, each cell's gene expression profile is written as a list of its most highly expressed genes (85).

**(iii) Biological foundation model as an encoder.** Instead of raw measurements, the foundation model's embeddings of a DNA or protein sequence are used as training data to fine-tune the LLM-based agent. The embeddings are also given to the LLM-based agent as input whenever it answers (19-21).

**(iv) Biological foundation model as a verifier.** During fine-tuning, a foundation model scores the LLM-based agent's answers, and the LLM-based agent learns to give answers that score well. Errors made by the foundation model are therefore learned by the LLM-based agent (86).

Only the encoder route gives the LLM-based agent a foundation model's representation of the sequence each time it answers. BioReason shows when this helps and when it does not. On questions that ask which disease a variant causes, given its gene and pathway described in text, the LLM alone reached 90% accuracy, and the encoder made little difference. On variant classification, which asks whether a variant is pathogenic from its DNA sequence, the LLM alone reached 49% accuracy, and adding Evo 2 raised this to 80%, compared with 61% for Nucleotide Transformer (21). When the answer lies within the LLM's textual knowledge, the encoder adds little; when it lies beyond, the encoder sets the ceiling of knowledge. Reinforcement learning improves how a system uses the knowledge it can reach but does not move the foundation model's boundary, and more reasoning cannot supply missing premises: a reasoning model's written steps are not guaranteed valid or faithful (87, 88), and biological inference, unlike mathematical proof (89), starts from incomplete premises that only experiments can test.

Access to a foundation model, however, does not ensure that an agent relies on it. Asked to interpret a noncoding variant at the 1p13 LDL cholesterol locus, the Paper2Agent named SORT1 as the causal gene, citing AlphaGenome's predicted expression score (78). SORT1 is indeed the experimentally established causal gene (90), but AlphaGenome scored CELSR2 and PSRC1 higher (23, 78). The agent also reported reduced SORT1 expression, whereas experiments show an increase (90). It thus reached the right gene through reasoning that contradicted both its own tool's output and the experimental evidence. An agent inherits a foundation model's knowledge boundary when it relies on the model's output, but a correct final answer cannot reveal whether it did.

## 4. KNOWLEDGE BOUNDARIES REVEALED BY EXISTING STUDIES

Both LLMs and biological foundation models have knowledge boundaries, but those of LLMs are rarely visible because modern language is densely represented in their training data. For biology, the decipherment analogy of Section 2.1.1 applies: decipherment asks whether any structure can be recovered, whether it carries meaning, whether it can be used to write, and which parts remain unreadable, and we use the same questions to organize the evidence below.

We focus on foundation models because, among the components an agent relies on, the boundaries of foundation models are the least characterized; an agent that calls a foundation model inherits that model's knowledge boundary. Agents with access to the same tools can differ in how well they select and combine them. We first discuss how each of the five design elements sets part of the boundary and then review what models have learned about biology and where that knowledge stops.

### 4.1. How the Five Design Elements Define the Knowledge Boundary

Each design element sets a different limit on what a model can learn and how it can be used (**Figure 2**). This holds for any model, including conventional machine-learning models; what is new is that foundation models are reused far beyond their original purpose, so these limits travel with them.

**Input data defines what can be known.** A model can learn only patterns that appear often in its training corpus. Human proteins and model organisms are represented by many related sequences and measured cell states, whereas poorly studied lineages and rarely profiled cell states appear seldom or not at all. Tokenization fixes the smallest unit the model can distinguish: a six-nucleotide token blurs a single-nucleotide variant, and a rank-based input discards how much each gene is expressed (Section 2.2.1).

**Model architecture sets which relationships can be captured efficiently.** A convolutional neural network detects short local patterns such as a transcription factor binding motif, whereas attention, the core of the Transformer, lets distant positions interact at a cost that grows roughly quadratically with input length (34). ChromBPNet, a convolutional model, reads about two kilobases of sequence at single-base resolution and is well suited to motifs and the variants within them (45). Enformer adds attention to integrate regulatory elements about 100 kilobases away (56), and HyenaDNA reaches one million nucleotides (35). An interaction longer than the input window cannot be learned, however much data is available.

**The training scheme defines what is rewarded and therefore what a model is good at.** ESM2's likelihoods score how well a mutation is tolerated (25), but it is not built to write new sequences, whereas ProGen2 and Evo are natural generators of new proteins and genomes (40, 41). SCimilarity's embeddings is well suited for searching an atlas for similar cells but not for predicting the response to a perturbation (43). It should be noted that none of these objectives of the models directly rewards learning cause and effect.

**Evaluation defines what becomes visible.** A model tested only on data resembling its training set can appear strong while failing on new proteins, species, or conditions; if test proteins have close relatives among the training proteins, a high score may reflect recall rather than learned rules. How a benchmark is split, and which simple baselines it includes, decides which weaknesses can be seen.

**Post-analysis sets what becomes accessible.** Probing, fine-tuning, and interpretability analysis each expose different knowledge. Knowledge may be present yet hard to extract, and a correct prediction does not show that the model has learned the relevant mechanism.

### 4.2. What Foundation Models Have Captured About Biology

Without labels, foundation models capture statistical patterns that reflect general properties of their building blocks, such as amino acid chemistry and which residues contact each other in a folded protein, and functional properties, such as which mutations a protein tolerates. They do not capture the physical mechanisms behind these patterns: a protein language model that predicts residue contacts has no representation of the physical and chemical forces that drive folding. To the decipherment questions we add whether knowledge extends beyond the training data and whether researchers can read it out. Each answer rests on comparing model outputs with independent data or on laboratory tests of designs, and each paragraph below is labeled with the design element that most directly shapes it.

**Data: can a model recover known biological patterns from unlabeled sequence?** For proteins, the evidence is strong. Protein language models are never told the properties of amino acids, yet their embeddings group residues by hydrophobicity and charge (91). They also recover which residues make physical interactions in the folded protein structure, because such residues tend to mutate together during evolution (91, 92). ESMFold builds on this to predict atomic-level structure from a single sequence (6), and AlphaFold, which also uses alignments of related sequences, reached near-experimental accuracy in the blind CASP14 assessment (4, 93).

**Training scheme: does a model capture what is functional?** The evidence is largely positive. Protein language model likelihoods predict mutational effects without training on mutational data and correlate with deep mutational scanning measurements (25). EVE, an unsupervised model of evolutionary variation, matched or exceeded supervised methods at classifying human variants as pathogenic or benign (59). On ProteinGym, a common benchmark of more than 200 deep mutational scanning assays, unsupervised models consistently tracked measured fitness, though only moderately, and no single model was best across all proteins (44).

**Training scheme: can a model design new, functional molecules?** Several studies show that it can. Design is a stricter test than scoring, and autoregressive training made it possible. ProGen produced artificial lysozymes with measurable catalytic activity (62), and ESM3, which can be prompted with structure and function tokens, generated a fluorescent protein distant from known natural sequences (30).

**Data: can a model generalize beyond the sequences and species it was trained on?** Generalization has improved, mainly through enriching what each input token encodes rather than through adding parameters. SaProt pairs each amino acid with a token for its local structure (52). GPN-MSA adds aligned sequences from 100 vertebrate genomes, making conservation visible, and outperforms sequence-only models at predicting deleterious variants (94). UCE represents each gene by the ESM2 embedding of its protein rather than by a fixed gene list, so it can place cells from species absent from training in a shared space (11).

**Post-analysis: can researchers read out what a model has learned?** Early studies suggest that they can. Sparse autoencoders (Section 2.1.1) have recovered features for active sites, zinc-finger domains, and disulfide bonds in protein language models (95-97). Model knowledge has also been used experimentally: simulated gene deletion with Geneformer nominated therapeutic targets for cardiomyopathy that were then supported experimentally (9), and general protein language models guided antibody evolution toward improved binding without task-specific training (98).

Together, these studies show that self-supervised training captures real biological knowledge, from residue chemistry to evolutionary constraint on function, that can be reused for variant interpretation and design. The clearest successes are in proteins, where the corpus is largest and the rules are written most directly into sequence.

### 4.3. Where Foundation Model Knowledge Stops

These successes raise the question of under what conditions predictions remain reliable. Studies addressing it, many conducted by researchers other than the models' developers, have compared protein, DNA, and single-cell foundation models with simpler alternatives, such as linear or logistic regression, one-hot encodings of sequence, or simply the average of the training data. We organize the limits these studies reveal by design element.

**Data: does a model learn biology, or the history of sampling?** In many cases it captures both, and the two cannot be readily distinguished. Protein language model likelihoods partly reflect how often related sequences were sampled across the tree of life rather than fitness alone (32). Gene function annotation can become circular: LLM-based annotators may score well on benchmarks partly because curated databases shaped both their training and the benchmark labels, and they perform worse on orphan genes and non-model organisms (12, 99). Whole data types remain thin: spatial, perturbation, and bulk corpora are small compared with sequence databases and single-cell atlases, and more parameters cannot recover patterns the data never contain.

**Architecture and scale: does a model trained on whole protein families detect the effect of one mutation?** In many cases, it does not. AlphaFold2 relies heavily on related sequences and largely fails to predict the structural disruption caused by destabilizing point mutations (100). Scale has not closed this gap: larger protein language models do not consistently improve zero-shot fitness prediction (44), and downstream tasks often reuse features learned early in pretraining (101). Input length sets a related limit. AlphaGenome reads one megabase of DNA at a time, an advance that underlies its state-of-the-art performance (23), yet regulation acting over longer distances or between chromosomes lies outside a single input.

**Training scheme: does a model trained on observed states predict the effect of a perturbation?** Current evidence suggests that it rarely outperforms simple baselines. Masked and autoregressive objectives reward predicting what occurs together, not what causes what. Enformer predicts regulatory signals across the genome well, yet differences in gene expression between individuals are not well explained and often predicts the wrong direction of effect (102, 103). A 2026 preprint reports the same gap for AlphaGenome and Borzoi (104). In predicting responses to genetic perturbations, simple linear models and the average of the training data match or outperform foundation models (105-107). Single-cell representations show the same pattern. Geneformer and scGPT embeddings did not outperform simple baselines without fine-tuning (108), an L1-regularized logistic regression rivaled scBERT on its reported task (109), parameter-free representations match single-cell foundation models (110), and GenePT suggests that part of what these models learn is already recorded in text (111). Similar results hold for mutation-dense protein design (112) and for multi-omics integration, where foundation models showed no uniform advantage across 23 methods (113).

**Evaluation: did a benchmark score reflect what a model can actually do?** In several cases, the score did not. DNA language models are spread across competing benchmarks, on which they showed inconsistent gains over lightweight supervised models (114, 115). Common perturbation metrics are dominated by systematic expression differences unrelated to the specific perturbation, so a model can score well simply by predicting an average response (116). Conversely, when metrics are calibrated against both an uninformative mean baseline and a positive control, several deep-learning perturbation models do outperform simple baselines (117), so part of the reported parity reflects the metric rather than the model.

The same gap between benchmark scores and real reliability appears when models face the data and decisions of practice. AlphaMissense showed discordance with curated clinical evidence across 5,845 missense variants in 59 disease genes (118). Structure-confidence scores poorly separated successes from failures among 614 experimentally characterized *de novo* protein designs (119). Most of 21 pretrained RNA language models did not outperform a one-hot encoding for secondary-structure prediction once closely related sequences were removed from the test set (120). In each case, a model that looked reliable by its usual measure was less reliable for its intended use.

**Post-analysis: is correct answer evidence that the model knows why?** A correct answer alone does not establish this. Fine-tuning can add knowledge from the new labels, so a strong fine-tuned score does not show how much was learned during pretraining. Attention weights should not be read as biological mechanisms, and sparse-autoencoder dictionaries so far cover only a few protein and genomic models (95-97). For most models, it remains unknown whether a correct prediction reflects real biology.

These studies shift the question from whether foundation models learn biology to what they have learned and where that knowledge stops. The boundary has several layers: knowledge missing because observations were never or rarely collected; knowledge hard to learn because the architecture cannot span it or the objective does not reward it, as with cause and effect; capability hidden or exaggerated by evaluation; and knowledge present in the model but inaccessible to current analysis. Benchmark performance, confidence scores, transferable embeddings, and reliability in application must therefore be established separately.

## 5. OPEN CHALLENGES: A BALANCED ASSESSMENT

The studies discussed above examine individual models. Here we step back to the field and assess three broader limitations: gaps in how models are evaluated, the assumption that one ever-larger general-purpose model will serve biology as it has served language, and the field's early developmental stage.

## 5.1. Gaps in How Foundation Models Are Evaluated

Independent evaluation of biological foundation models is concentrated in a few areas, mainly protein sequence, single-cell representation, and perturbation prediction, and usually tests only one design element, the evaluation scheme.

**Uneven evaluation across modalities and tasks.** Bulk transcriptomics has no dedicated foundation-model benchmark, spatial benchmarks are emerging, and multi-omics and RNA models have few head-to-head comparisons (113, 120). Protein structure, protein fitness, and perturbation prediction have blind or community-maintained evaluations (4, 44, 121), whereas sequence-to-function models are evaluated mainly on developer-defined suites, with cross-individual tests appearing only recently (102-104). Few models report whether they can signal that a query lies outside their boundary.

**Agents and LLMs as users of foundation models.** Agent benchmarks tend to test general bioinformatics analysis (15, 16), not whether an agent chooses the right foundation model or interprets its output correctly. For systems that feed foundation-model embeddings into an LLM (21, 122), controls that separate the encoder's contribution from the LLM's contribution are rarely reported; when such controls were applied to six systems, BioReason and BioReason-Pro were as accurate with shuffled biological inputs as with intact ones (123). Evidence on adoption by biologists remains limited to surveys (124, 125).

## 5.2. Betting on a Single General-Purpose Model

Much of the field's momentum comes from an analogy with LLMs, whose performance improves predictably as parameters, training data, and computing budget grow, a relationship known as scaling laws (126). Biological foundation models have followed a similar trajectory, yet larger protein language models do not consistently predict mutational effects better than their predecessors (44, 101). Language has a comparatively unified structure, with a shared grammar and stable vocabulary, and this supports a single general-purpose model. Biology is heterogeneous across scale, from molecule to cell to tissue to organism, and across data type, species, and experimental condition. The language analogy therefore only partially holds: biology's vocabulary is not fixed, its grammar is largely unknown, and its corpus is incomplete. Multimodal proposals attempt to bridge these differences (37, 55), but so far each bridges only a subset of data types. Unlike web-scale text, biology's corpus is still growing as new technologies produce new kinds of data. The LLM field's development trajectory is therefore a useful comparison but not a reliable forecast: biological foundation models may improve more through new data types and designs matched to them than through scale alone.

**5.3. AI for Biology Is at an Early Stage of Development**

AI for biology is an emerging discipline in its infancy. Positioning the field on a developmental trajectory lets us understand its present strengths and limitations as features of a transient stage rather than as a fixed outcome. We consider two complementary lenses. The first lens is scientific maturity. Biology has historically progressed from observation and classification, as in Darwin's cataloguing of species, to mechanistic understanding of structure, function, and causality, and ultimately to engineering: the design of biological systems from first principles. Much of the foundation-model work surveyed here sits at the first stage, offering pattern-level prediction from observational data, and the repeated finding that simple baselines match foundation models is consistent with this positioning. Descriptive advances are necessary precursors to mechanistic understanding, but claims framed in engineering-stage language, such as *de novo* design, should be read with this stage in mind. The second lens is evolutionary competition. Periods of open ecological opportunity have historically favored large, generalist strategies, whereas the adaptive radiation that follows favors smaller organisms fitted to specific niches (**Figure 3B**). The current expansion of biological foundation models resembles the first, open-niche phase. Whether the field moves toward specialized models that fill the boundary regions where general models perform poorly (**Figure 3C**) is a consequential consideration in setting resources and research priorities.

## 6. FUTURE DIRECTIONS AND RECOMMENDATIONS

### 6.1. Specialized Models, Coordinated by Agents

Because biological data are so heterogeneous and sparse, the field need not wait for a single general-purpose model. The evidence reviewed here points instead toward a portfolio of broad models that supply transferable embeddings, alongside narrower models fitted to the boundary regions where general models perform poorly (**Figure 3C**). GPN-MSA, which draws on evolutionary alignments (94), and ChromBPNet, which targets a single assay at base resolution (45), already follow this pattern. Agents are well placed to manage such a portfolio. They can route each question to the model whose boundary covers it, combine outputs from complementary models, and fall back to a simple baseline when no model applies.

### 6.2. Making Knowledge Boundaries Explicit to Agents

For an agent to route each question to the right model, it must know where each model's boundary lies. Current tool descriptions state what a tool does but seldom where it fails, so errors from outside a model's boundary can propagate silently: a variant score from a model trained without population data looks as authoritative as a well-supported result, and recent high-profile agent studies still rely on human scientists to run experiments and interpret results (82, 83, 127). We propose that each foundation model be released with a structured boundary statement, in the spirit of model cards and datasheets (128, 129). Organized by the five design elements, it would record the species, assays, and conditions in the training data; the context length and tokenization of the architecture; the training objective; the benchmarks used, including comparisons with simple baselines; and the post-analysis methods under which the reported results hold. In machine-readable form, the statement would help a biologist decide whether to trust a result and an agent to decide whether to call the model. Just as GeneAgent checks its claims against curated databases (12), an agent could confirm that a query lies within a model's training and evaluation domain before relying on its output. Boundary statements would also show where new data are most needed. Self-driving laboratories, in which robotic platforms run experiments chosen by a model and return the results for the next round, can target these regions directly (130, 131); combined with agents that plan experiments and foundation models that propose candidates, such closed loops could extend a model's knowledge boundary with new measurements rather than more of the same data. The knowledge boundary would then become the interface between foundation models and agentic systems.

### 6.3. Community as Scientific Infrastructure

Making knowledge boundaries explicit is also a problem of community organization. Open science offers precedents. UniProt and the Protein Data Bank became trusted infrastructure (2, 28), open-source ecosystems made models widely reusable (132), and community assessments such as CASP and CAGI established credible measures of progress (93, 133). Embedded scientist programs, reciprocal laboratory placements, fellowships, and professional community-management roles offer deeper cross-disciplinary exposure than short courses (134). The aim is shared literacy rather than making every biologist a model builder. A community-maintained registry linking each model to its boundary statement, independent evaluations, known limitations, and replication results (135, 136) would give biologists and agents a living record of each model's knowledge boundary.

## 7. CONCLUSION

Artificial intelligence for biology is advancing along two complementary paths. Foundation models compress large biological datasets into reusable representations, and agentic systems compose existing literature, databases, and software into automated workflows. Agents are bound by what has already been documented, and foundation models are bounded by the data, architecture, training scheme, evaluation scheme, and post-analysis that define them. We present our perspective that integrating the two paths depends on making the knowledge boundary of foundation models explicit. The five design elements provide a common template for describing any model; its inputs and outputs locate it within the landscape, and published evidence on what models capture and miss shows where current knowledge boundaries lie. Making these boundaries explicit gives biologists a principled basis for deciding when to trust a model, gives model builders a concrete target for improvement, and gives agents the information they need to select and interpret models appropriately. As biological measurement continues to expand, clearly stated knowledge boundaries can serve as a shared language between AI and biology.

## DISCLOSURE STATEMENT

The authors are not aware of any affiliations, memberships, funding, or financial holdings that might be perceived as affecting the objectivity of this review.

## ACKNOWLEDGMENTS

AI assistance (Claude, Anthropic, ChatGPT) was used for literature collection, fact verification, and language editing; the key messages, content structure, manuscript framework, and figures are entirely the authors' own, and the authors take full responsibility for the content.

## AUTHOR BIOGRAPHIES

This review is written by a group whose individual paths span the divide it describes. Tian Cai has a background in computer science and deep learning model design and joined St. Jude Children's Research Hospital for immersive, hands-on exposure to biological research. Andrew Kleist is a structural biologist and physician. Jaimin Patel is a principal software engineer who adapts industry technologies for academic research. Alex Taylor recently completed a degree in computer science and is now building expertise in biology. Malvika Sharan is a community builder in open science, and Renato Umeton works in the space of AI and agentic systems. Duccio Malinverni is a physicist interested in biology. M. Madan Babu works at the interface of data science and biology and is among the scientists who bridge the two communities.

## DATA AND CODE AVAILABILITY

The online resource accompanying this review, an interactive landscape of the foundation models, agentic systems, concepts, and techniques discussed here, is available as a HTML file and at [**URL**]. Code is released under the MIT License and content under [CC BY 4.0].